\documentstyle[a4wide,12pt]{article}
\title{On the quantization of angular momentum}
\author{Vu B Ho\\Department of Physics\\Monash University\\
Clayton Victoria 3168\\Australia}
\date{}
\begin{document}
\maketitle
\begin{abstract}
It is shown that when a hydrogen-like atom is treated as a two dimensional
system whose configuration space is multiply connected, then in order to
obtain the same energy spectrum as in the Bohr model the angular momentum
must be half-integral.
\end{abstract}

\newpage
In physics singular behaviours of a physical quantity are sometimes due to
a breakdown of the coordinates used for its description, and in such cases
singularities can be removed by suitable coordinate transformations. A well
known such singularity is the Schwarzschild singularity at $r=2m$ in general
relativity. In Einstein's theory of general relativity, the Schwarzschild
metric takes the following form \cite{Mis}

\begin{equation}
ds^2=-\left(1-\frac{2m}{r}\right)dt^2+\left(1-\frac{2m}{r}\right)^{-1}dr^2
+r^2\left(d\theta^2+sin^2\theta d\phi^2\right).
\end{equation}
It is clear that with solution in this form the metric components become
singular at both $r=2m$ and $r=0$. However it has been shown that while the
singularity at $r=0$ is a true physical singularity, the singularity at
$r=2m$ is only a coordinate breakdown \cite{Kru,Wald}. (For more general
discussions on coordinate transformations to eliminate Schwarzschild
singularities, see \cite{Ho}.)

In this letter we consider coordinate transformations in quantum mechanics
and the problem of quantization of angular momentum in multiply connected
spaces. In quantum mechanics the problem of multivalued eigenfunctions of
angular momentum may be argued to appear only because one changes from
Cartesian coordinates to polar coordinates which are singular at the
coordinate origin \cite{Merz}. On the other hand, recent developments of
quantum mechanics in multiply connected spaces have shown that, within the
present physical interpretation of quantum mechanics, the use of multivalued
wave functions is allowable provided the configuration space is nonsimply
connected \cite{Mor,Bala}. In the following we will discuss the particular
case of a hydrogen-like atom from the point of view of an observer who is in
a coordinate system that describes the atom as a planar physical system. In
this case if we assume that the electron can never be able to penetrate the
nucleus, then the configuration space can be considered as multiply
connected and the use of multivalued wavefunctions is allowed.
In such situation the single-valuedness condition is no longer obvious to be
a good requirement for the angular momentum to be integral. However we will
show that with this observer the eigenvalues of angular momentum must be
half-integral if the observer obtains the same spectrum of energy as that
of an observer who observes the atom in three dimensional space.

The hydrogen-like atom consisting of a single electron of charge $-e$ and a
nucleus of charge $Ze$ is described by the eigenvalue equation

\begin{equation}
-\frac{\hbar^2}{2m}\nabla^2\psi({\bf r}) - \frac{Ze^2}{r}\psi({\bf r}) =
E\psi({\bf r}),
\end{equation}
where $m$ is the reduced mass. We will consider the case of bound
states with $E<0$. In three dimensional space, when the wave function is
written in the form

\begin{equation}
\psi({\bf r}) = R(r)Y_{l,m}(\theta,\phi),
\end{equation}
we get the radial equation for the function $R(r)$ \cite{Lan,Bra}

\begin{equation}
\frac{d^2R}{d\rho^2}+\frac{2}{\rho}\frac{dR}{d\rho}-\frac{l(l+1)}{\rho^2}R
+ \frac{\lambda}{\rho}R - \frac{1}{4}R = 0,
\end{equation}
where $\rho$ and $\lambda$ are defined below

\begin{equation}
\rho=\left[\frac{8m(-E)}{\hbar^2}\right]^{1/2}r, \ \ \ \ \
\lambda=\left[\frac{Z^2e^4m}{2\hbar^2(-E)}\right]^{1/2}.
\end{equation}
$Y_{l,m}(\theta,\phi)$ are the familiar spherical harmonics which are
simultaneous eigenfunctions of ${\bf L}^2$ and $L_z$. By the
single-valuedness requirement, the quantum number $m$ must be integral.
On the other hand, the integral-valuedness of the quantum number $l$ is
required for the consistency of group representation \cite{Merz}. In fact
in the case of hydrogen-like atoms, the quantum number $l$ must be integral
if we want to obtain the same energy spectrum of the Bohr model. When we
seek solutions for $R(r)$ in the form

\begin{equation}
R(r)=\exp(-\rho/2)\rho^lS(\rho),
\end{equation}
then by substitution into the equation (4) we obtain the following
differential equation for the function $S(\rho)$

\begin{equation}
\frac{d^2S}{d\rho^2} + \left(\frac{2l+1}{\rho}-1\right)\frac{dS}{d\rho} +
\frac{\lambda-l-1}{\rho}S=0.
\end{equation}
This equation can be solved by a series expansion of $S(\rho)$

\begin{equation}
S(\rho)=\sum_{n=0}^\infty a_n\rho^n,
\end{equation}
with the coefficients $a_n$ satisfy the recursion relation

\begin{equation}
a_{n+1}=\frac{n+l+1-\lambda}{(n+1)(n+2l+2)}a_n.
\end{equation}
This result and the relation (5) show that the quantum number $l$ must be
integral for the energy $E$ to have the same form as that of the Bohr model.

Now let us examine the hydrogen-like atom with the view point of an
observer who sees it as a planar system. Since the Schrodinger equation of
form (2) is invariant under rotations, this observer can still invoke the
Schrodinger equation for an analysis of the dynamics of the atomic system.
However in this case there are two important aspects which relate to the
topology of the system we must emphasize and exploit in our discussion.
First the configuration space of the atom now is multiply connected, and
as mentioned above multivalued wave functions can be used. Second in two
dimensional space the more suitable coordinate system is the planar polar
coordinates, and so the Schrodinger equation now takes the form

\begin{equation}
-\frac{\hbar^2}{2m}\left[\frac{1}{r}\frac{\partial}{\partial r} \left(r
\frac{\partial}{\partial r}\right) + \frac{1}{r^2}
\frac{\partial^2}{\partial \phi^2}\right]\psi(r,\phi)-
\frac{Ze^2}{r}\psi(r,\phi) = E\psi(r,\phi),
\end{equation}
here we have assumed that the Coulomb potential remains its form. Solutions
of the form $\psi(r,\phi)=R(r)\Phi(\phi)$ then reduce the above equation to
two separate equations for the functions $\Phi$ and $R$

\begin{eqnarray}
\frac{d^2\Phi}{d\phi^2}+\mu^2\Phi&=&0,\\
\frac{d^2R}{dr^2}+\frac{1}{r}\frac{dR}{dr} - \frac{\mu^2}{r^2}R +
\frac{2m}{\hbar^2}\left(\frac{Ze^2}{r}-E\right)R&=&0,
\end{eqnarray}
where $\mu$ is identified with the angular momentum of the system. From the
equation (11) we obtain a solution for the function $\Phi$ in the form

\begin{equation}
\Phi(\phi) = \exp(i\mu\phi).
\end{equation}
With this form of solutions we normally require the angular momentum $\mu$
take integral values so that the single-valuedness condition is satisfied.
However, we will show that the integral-valuedness requirement for the
quantity $\mu$ in this case is not compatible with the assumption that an
observer in two dimensional space also obtain a spectrum of energy like
that of the Bohr model. It should be emphasized here that the Bohr model is
also a planar system.

{}From the equation (12), if we define $\rho$ and $\lambda$ as in (5) then we
can put this equation in a simpler form

\begin{equation}
\frac{d^2R}{d\rho^2} + \frac{1}{\rho}\frac{dR}{d\rho} -
\frac{\mu^2}{\rho^2}R + \frac{\lambda}{\rho}R - \frac{1}{4}R=0.
\end{equation}
We notice the crucial difference between the equation for $R$ in this case
and the equation (4) is the factor 2 before $dR/dr$. This equation can also
be solved by writting $R$ in the form (6) as before. The equation for the
function $S(\rho)$ now is slightly different from the equation (7) by the
coefficient before $S$

\begin{equation}
\frac{d^2S}{d\rho^2}+\left(\frac{2\mu+1}{\rho}-1\right)\frac{dS}{d\rho} +
\left(\frac{\lambda-\mu-1/2}{\rho}\right)S=0.
\end{equation}
A power series solution (8) for the function $S$ will result in the following
recursion relation

\begin{equation}
a_{n+1}=\frac{n+\mu+1/2-\lambda}{(n+1)(n+2\mu+1)}a_n.
\end{equation}
With this result and by the relation (5), the energy spectrum in this case
can be written explicitly in the form

\begin{equation}
E=-\frac{Z^2e^4m}{2\hbar^2(n+\mu+\frac{1}{2})^2}.
\end{equation}
Hence if the hydrogen-like atom is viewed as a two dimensional physical
system and if the energy is observed to have the same spectrum as that of
the Bohr model then the angular momentum $\mu$ must take half-integral
values. However it can be verified that integral values for the angular
momentum $\mu$ can be retained if we add to the Coulomb potential a
quantity -$\left(\hbar\sqrt{E/2m}\right)/r$ when the hydrogen-like atom is
viewed as a two dimensional physical system.

I would like to acknowledge the financial support of APA Research Award.

\end{document}